\documentclass[aps,prb,twocolumn,superscriptaddress,floatfix]{revtex4-1}
\usepackage[utf8]{inputenc}

\usepackage{graphicx}
\DeclareGraphicsExtensions{.png,.jpg,.pdf}

\usepackage{xcolor}
\usepackage{amsmath}
\usepackage{amssymb}

\usepackage{dcolumn}
\usepackage{bm}
\usepackage{hyperref}
\hypersetup{colorlinks=true,linkcolor=blue,citecolor=blue}
\usepackage{float}
\usepackage{siunitx}

\begin{document}

\title{Designer fermion models in functionalized graphene bilayers}

\author{N. A. García-Martínez}
\affiliation{Departamento de Física Aplicada, Universidad de Alicante, 03690 Spain}
\author{J. Fernández-Rossier \footnote{On leave from Departamento de Física Aplicada, Universidad de Alicante,  Spain} }
\affiliation{International Iberian Nanotechnology Laboratory (INL),
Av. Mestre José Veiga, 4715-330 Braga, Portugal}

\date{\today}

\begin{abstract}
We propose a method to realize a broad class of tunable fermionic Hamiltonians in graphene bilayer. 
For that matter, we consider graphene bilayer functionalized with sp$^3$ defects that induce zero energy resonances hosting an individual electron each. The application of an off-plane electric field opens up a gap, so that the zero energy resonance becomes an in-gap bound state whose confinement scales inversely with the gap.
Controlling both the distance among the defects and the applied electric field, we can define fermionic models, even lattices, whose hoppings and Coulomb interactions can be tuned.
We consider in detail the case of triangular and honeycomb artificial lattices and we show how, for a given arrangement of the sp$^3$ centers, these lattices can undergo an electrically controlled transition between the weak and strong coupling regimes.
\end{abstract}

\maketitle

The amazingly rich panoply of electronic phases that occur in solid state matter emerges from the interplay of kinetic energy of the electrons and the Coulomb interaction, both electron-electron and electron-ion. These energy scales are defined by the chemical composition and the crystal structure of each material. Physicists have been looking for strategies to create artificial lattices to confine fermions where the geometry and the energy scales can be tuned independently in order to explore emergent electronic phases escaping the dictatorship of chemistry.
Examples of artificial lattices include optical traps for cold atoms\cite{Cocchi2016}, arrays of quantum dots\cite{mortemousque2018,dehollain2019}, and surface adatoms\cite{gomes2012,drost2017,girovsky2017,khajetoorians2019}.

None of these approaches has reached yet the point where a systematic exploration of large scale lattices with non-trivial phases can be controlled in the quantum regime.
Optical traps can create artificial lattices with one fermion per site, but the hopping and Coulomb repulsion energy scales are not much larger than temperature,\cite{Cocchi2016} a requirement to observe non-trivial quantum behaviour.
Gate defined quantum dot arrays can definitely avoid that problem\cite{dehollain2019}, but the largest quantum dot array reported so far has only 9 quantum dots\cite{mortemousque2018}.
\emph{Scanning Tunnelling Microscope} (STM) has been used to arrange hundreds of carbon monoxide molecules\cite{gomes2012} and thousands of chlorine atoms\cite{Kalff2016} on a copper surface, defining artificial lattices for electrons in surface states. However, this approach does not allow further electrical control of the carrier density and the electronic states have finite lifetimes on account of their coupling to the substrate. 

\begin{figure}[h!]
\centering
 \includegraphics[width=0.5\textwidth]{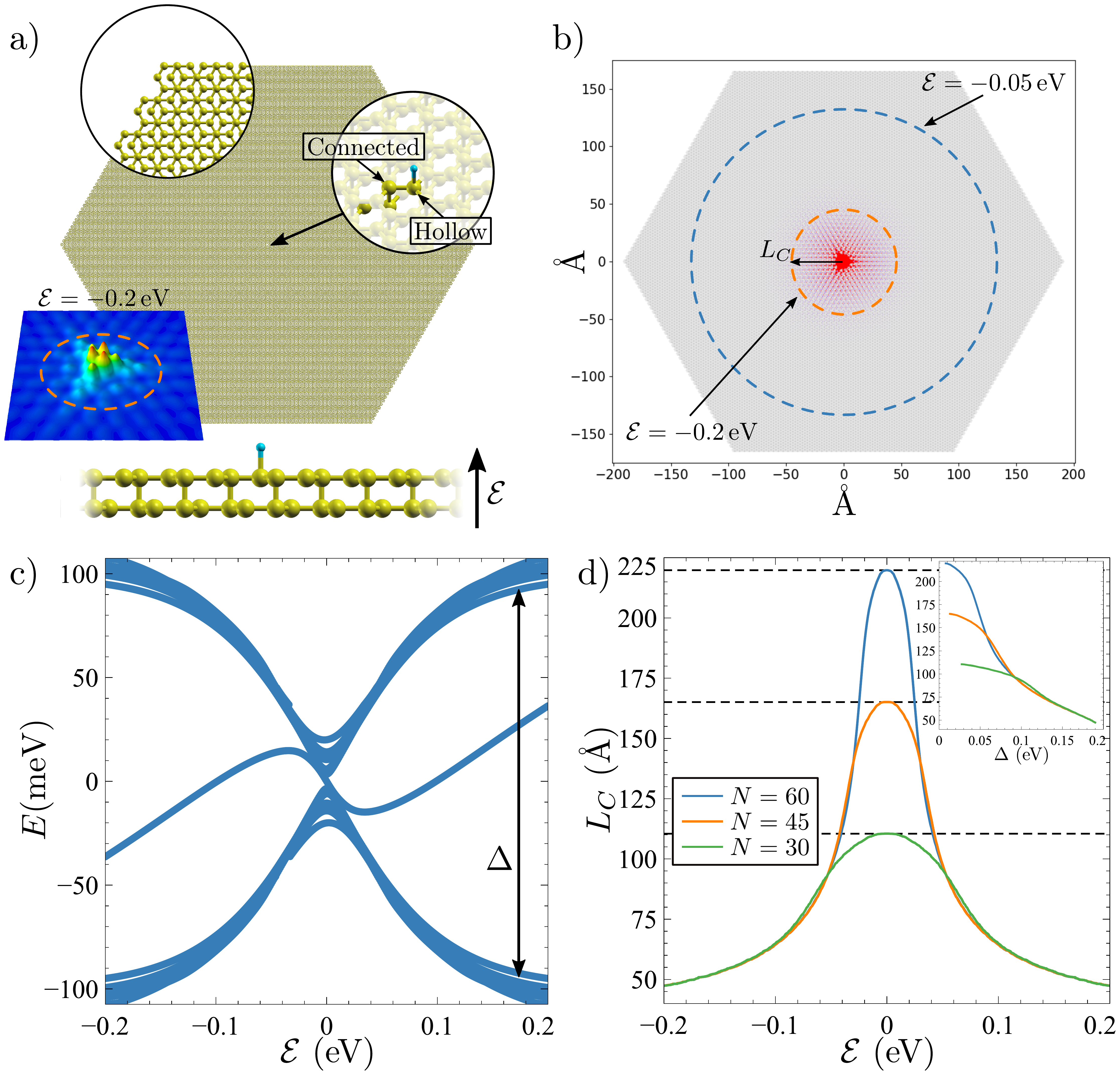}
\vspace{-5pt}
\caption{a) A bilayer graphene flake with armchair edges and a chemisorbed H adatom in the presence of an electric field is suggested as a platform to realize a number of fermionic models. The building blocks of these models are the electronic states localized around the sp$^3$ as shown in the 3-D simulation of the local density of states. b) Spatial distribution of the in-gap state created by a $sp^3$ defect in bilayer graphene in the presence of an electric field $\mathcal{E}=\SI{-0.2}{\eV}$. The two dashed lines show the confinement length for two values of the electric potential difference. c) Evolution of the 12 eigenenergies closest to $E=0$ with the electric field. d) Confinement length $L_C$ as a function of the electric field for different size islands. The dashed horizontal lines mark the maximal $L_C$ as given by the island size.}
\label{figure1}
\end{figure}

Here we propose functionalized graphene bilayer with an electrically controlled band gap as a flexible and tunable platform to define both fermion and spin model Hamiltonians with a variety of 1D and 2D lattices with a highly tunable energy scales and filling factors.
The building block of the model Hamiltonians are in-gap localized states, produced by sp$^3$ functionalization of carbon atoms in hollow sites (see Fig~\ref{figure1} a)). This can be achieved by atomic hydrogen chemisorption\cite{Brihuega2016}, but there are many other molecules that produce the same effect\cite{Santos2012}.
Arrays of these point defects can be created with state of the art STM manipulation,\cite{Brihuega2016} but self-assembly of sp$^3$ molecular centers could be envisioned.
Our calculations show that this platform permits to emulate a number of emblematic lattices (triangular, honeycomb, Kagome, rectangular)
with tunable exchange and Coulomb energies, that implement Hubbard and Heisenberg interactions. 
We show that artificial lattices which realize model Hamiltonians can be created in graphene bilayer which allows the exploration of regimes that lead to the emergence of 2D spin-liquid phases, Mott transitions, quantized anomalous Hall phases, and fractionalized Haldane spin chains and correlated superconductivity. 
 
Different aspects of our proposal have been independently verified experimentally. First, several experiments have demonstrated that an electric field opens up a gap as large as $\SI{250}{meV}$ in graphene bilayer\cite{Castro2007, Zhang2009, Taychatanapat2010}. Second, the generation of localized zero modes by sp$^3$ functionalization has been demonstrated by hydrogen chemisorption in graphene\cite{Brihuega2016}, together with the lateral manipulation of the hydrogen atoms, with atomic scale resolution\cite{Ugeda2011,Brihuega2016}. Last, but not least, the observation of correlated electronic phases in twisted graphene bilayer\cite{cao2018a, cao2018b}, 
that is believed to be associated to the formation of arrays of localized states in a moiré pattern\cite{}, shows that graphene bilayers can indeed host strongly correlated phases.
Here we propose to induce an array of states by chemical functionalization, which would permit to control the location of the bounded states. 
 
We consider Bernal stacked graphene bilayers, shown in Fig.~\ref{figure1} a). We use the standard single orbital tight binding model\cite{McCann2012} and consider first neighbour intralayer hopping $t=\SI{-2.7}{\eV}$ and first neighbour interlayer hopping $\gamma=\SI{0.4}{\eV}$ in accordance with the literature\cite{KatsnelsonBook}.

The effect of sp$^3$ functionalization is included by removal of a site in the lattice\cite{}: the $p_z$ orbital of the carbon atom forms a strong covalent bond with the functional group and is effectively removed from the $p_z$ lattice\cite{}. We revisit first the problem of a single sp$^3$ defect in graphene bilayer\cite{Castro2010}, focusing on the influence of an off-plane electric potential difference, denoted $\mathcal{E}$, which has been considered already in the literature\cite{Nilsson2007}.
In the upper layer of Bernal stacked bilayer graphene one of the sublattices is connected to the lower layer while the other sublattice lay on {\it hollow} positions. We will consider only functionalization of atoms in hollow sites (see Fig~\ref{figure1} a)). 

At zero electric field, $\mathcal{E}=0$, pristine graphene bilayer is a zero gap semiconductor, with a degenerate conduction and valence band at the $K$ and $K'$ points of the Brillouin zone\cite{McCann2012}. The sp$^3$ functionalization creates a zero mode resonance \cite{Nilsson2007,Castro2010}, as expected in any bipartite lattice with a gapless spectrum and with a missing site.\cite{Lieb1989} 
The application of an electric field opens up a gap in the pristine bilayer spectrum and, as we show below, the zero energy resonance becomes a bound state.

In order to study the single particle spectra of graphene bilayer with sp$^3$ functionalizations we resort to Lanczos diagonalization\cite{Lanczos1950, Arnoldi1951} of the single-particle tight-binding model, due to the very large number of carbon sites. For instance, the island used in the calculations of Fig.~\ref{figure1} has $N=131772$ sites and a corresponding quantum confinement gap of $\sim\SI{7}{\meV}$ which becomes irrelevant when the gap induced by the electric field is much larger. 
The results of Fig.~\ref{figure1} c) show the evolution of the in-gap state as well as a few conduction and valence states as a function of the off-plane electric field. It is apparent that as the field is increased, a gap opens and the zero mode resonance stays inside the gap, becoming a bound state with a normalized wave function.


Importantly, the extension of the in-gap bound state induced by the sp$^3$ functionalization can be electrically tuned. In Fig.~\ref{figure1} d) we show the confinement length of the in-gap state as a function of the electric field calculated as the minimal radius which contains $> 98\%$ of the in-gap wave function.
The extension of the in-gap state is controlled by the size of the energy gap, as shown in the inset of Fig.~\ref{figure1} d). This result that can be obtained analytically as well\cite{Nilsson2007}. This is an important resource in order to control the extension of the in-gap states, and hence the overlap among defects. This will be one of the key ingredients to control the energy bandwidth in artificial lattices and the effective Hubbard repulsion.

\begin{figure}[h!]
\centering
\includegraphics[width=0.5\textwidth]{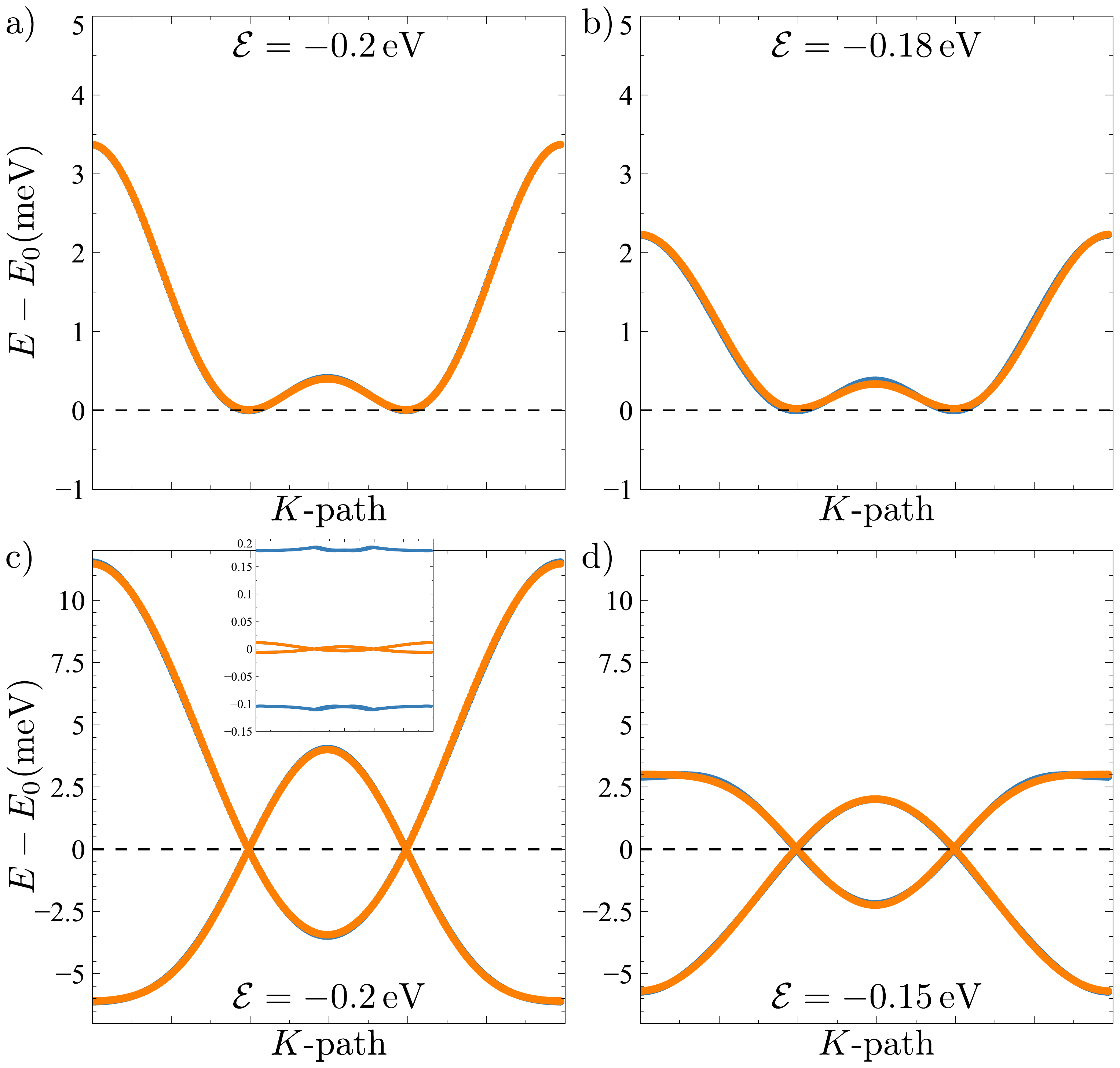}
\vspace{-15pt}
\caption{In-gap energy bands for a triangular (a,b) and a honeycomb lattice (c,d) of sp$^3$ defects on a graphene bilayer for two different values of the electric field. The inset of panel c) also shows the conduction and valence bands. The bands obtained by fitting to tight-binding models for triangular and honeycomb lattices with one orbital per site and up to third neighbour hopping are also plotted in blue, but barely visible due to the perfect fitting.}
\label{figure2}
\end{figure}

We now study the single particle states of artificial arrays of sp$^3$ functionalization. We have considered two geometries: triangular and honeycomb, although many other structures, such as Kagome and rectangular, are possible by functionalization of the hollow sublattice of the top layer. The calculation is carried out using a supercell
with one (triangular) or two (honeycomb) sp$^3$ defects. As expected, we obtain as many in-gap bands as sp$^3$ defects in the unit cell. The corresponding in-gap energy bands for triangular and honeycomb artificial lattices with different electric fields are shown in Fig.~\ref{figure2}, for two different values of the off-plane electric field.

The in-gap bands can be perfectly fitted to tight-binding models for the triangular and honeycomb lattice with hopping up to third neighbours. These energy bands arise because, in the presence of an electric field, the sp$^3$ in-gap states are no longer sublattice polarized. Therefore, adjacent in-gap states can hybridize, leading to the formation of in-gap bands. 
The size of the unit cell determines the distance between the point defects and the electric field determines the extension of the bound states.
Thus, both the density of defects and the electric field determine the bandwidth which is in the range of $\SI{10}{\meV}$ for the cases shown in Fig.~\ref{figure2}.
 
The results of Figs.~\ref{figure1} and \ref{figure2} show that functionalization of gapped graphene bilayer with ordered arrays of sp$^3$ defects lead to the formation of in-gap energy bands that can be understood in terms of a tight-binding lattice model, where every functionalization corresponds to a site and a relatively short ranged hopping matrix. 
Importantly, the density of carriers can be tuned independently of the size of the graphene bilayer, using dual gating\cite{Zhang2009,Taychatanapat2010}.Therefore, the Fermi energy could be moved along the bandwidth, controlling the filling factor, very much like in the case of the flat bands in the magic angle twisted bilayer\cite{cao2018a,cao2018b}. 

We now consider the effect of electron-electron interactions. For that matter, we model the Coulomb interaction for the functionalized graphene bilayer in the Hubbard approximation. Our main goal is to provide a fair estimate
of the magnitude of the effective Coulomb repulsion for the electrons occupying the in-gap bands of the artificial lattices. We consider the exact solution for the minimal system where hybridization and Coulomb repulsion compete with each other, namely, a dimer at half filling. The energy scales of this problem are controlled by both the distance between the two sp$^3$ defects and the gap induced by the electric field.

We first solve the single particle problem for two impurities, using again a bilayer island (with up to N=131772 sites) and two sp$^3$ centers. The diagonalization yields two in-gap states, with energy $\epsilon\pm \frac{\delta}{2}$, where $\delta$ is the hybridization splitting of the states. 
Two states $\phi_{1/2}$, localized states around the defects, can be built out of the two in-gap eigenstates, $\psi_\pm$:
\begin{equation}
  |\phi_{1/2}\rangle = \frac{1}{\sqrt{2}}\left(|\psi_+\rangle \pm|\psi_-\rangle\right)
\end{equation}
In the following we refer to $\phi_1$ and $\phi_2$ as site states, but it has to be kept in mind that they are extended states living in hundreds of carbon atoms. 

We now build the many body Hamiltonian for two electrons in two states, $\phi_{1,2}$. The Hamiltonian can be written as the sum of the singe particle part $\mathcal{H}_0$ and the Coulomb interaction part $\mathcal{H}_{U}$. The first term reads:
\begin{equation}
{\cal H}_0= \tilde{t} \sum_{\sigma} \left(c^{\dagger}_{1\sigma}c_{2\sigma}+c^{\dagger}_{2\sigma}c_{1\sigma}\right)
\end{equation} 
where $c^{\dagger}_{\eta,\sigma}$, with $\eta=\{1,2\}$ creates an electron with spin $\sigma$ in the site state $\phi_{\eta}$ and $\tilde{t}=\frac{\delta}{2}$

A minimal model for the Coulomb contribution is obtained using the Hubbard approximation and truncating the Hilbert space to just the two in-gap states. When represented in the basis of the two site-orbitals, the Hubbard term can be expressed as the sum of four types of contributions\cite{Ortiz2019}:
\begin{eqnarray}
{\cal H}_{ U}&=& 
\left(\tilde{U}_1-\frac{J}{4}\right) n_{1\uparrow} n_{1\downarrow} +
\left(\tilde{U}_2 -\frac{J}{4}\right)
 n_{2\uparrow} n_{2\downarrow} 
 -\nonumber\\
&-& J \vec{S}_1\cdot\vec{S}_2+ {\cal V}_{\rm pair}+{\cal V}_{12}+ {\cal V}_{21}
\label{HAMIL}
\end{eqnarray}
where $n_{\eta\sigma}= c^{\dagger}_{\eta\sigma}c_{\eta\sigma}$ is the occupation operator in the basis of $\phi_{1,2}$, 
$\vec{S}_\eta= \frac{1}{2} \sum_{\sigma,\sigma'} c^{\dagger}_{\eta\sigma}\vec{\tau}_{\sigma,\sigma'}c_{\eta\sigma'}$ are the spin operators associated to these quantum states.

Hubbard-like terms, with energy $\tilde{U}_{\eta}-\frac{J}{4}$ describe the Coulomb energy penalty of double occupation of states $\phi_1$ and $\phi_2$. The energy scales $\tilde{U}_{\eta}$ are given by 
$\tilde{U}_{\eta}= U \sum_i |\phi_{\eta}(i)|^4$
where $\sum_i |\phi_{\eta}(i)|^4$ is the so-called inverse participation ratio (IPR) and it is a metric of the extension of the state\cite{Ortiz2019} and $U$ is the atomic carbon on-site repulsion. Because of the symmetry of site states, we have $\mathcal{U}_1\simeq\mathcal{U}_2$.
Here we take $U=|t|=\SI{2.7}{\eV}$. The second type of term is a ferromagnetic exchange
with 
$J= 2U \sum_i |\phi_{1}(i)|^2|\phi_{2}(i)|^2$
which is a metric of the \emph{overlap} of the two site states. 
The last two terms are \emph{density assisted} hopping terms, 
\begin{equation}
  \mathcal{V}_{21} = \sum_{\sigma} n_{1\sigma}
\left(
t_{12} c^{\dagger}_{1\overline{\sigma}} c_{2\overline{\sigma}}
+t_{12}^* c^{\dagger}_{2\overline{\sigma}} c_{1\overline{\sigma}} \right)
\end{equation}
and the pairing hopping term:
\begin{equation}
\mathcal{V}_{\rm pair}=
\Delta c^{\dagger}_{1\uparrow} c^{\dagger}_{1\downarrow} c_{2\uparrow}c_{2\downarrow}
+ \Delta^* c^{\dagger}_{2\uparrow} c^{\dagger}_{2\downarrow} c_{1\uparrow}c_{1\downarrow} 
\end{equation}

The Hubbard matrix elements that control these Coulomb energies assisted with hopping processes are: 
$t_{12}= U \sum_i |\phi_{1}(i)|^2 \phi_1(i)^* \phi_{2}(i)$
and
$\Delta= U \sum_i (\phi_{1}(i)^*)^2 \phi_2(i)^2$.
Both $t_{12}$ and $\Delta$ could be complex numbers.

\begin{figure}[h!]
\centering
\includegraphics[width=0.5\textwidth]{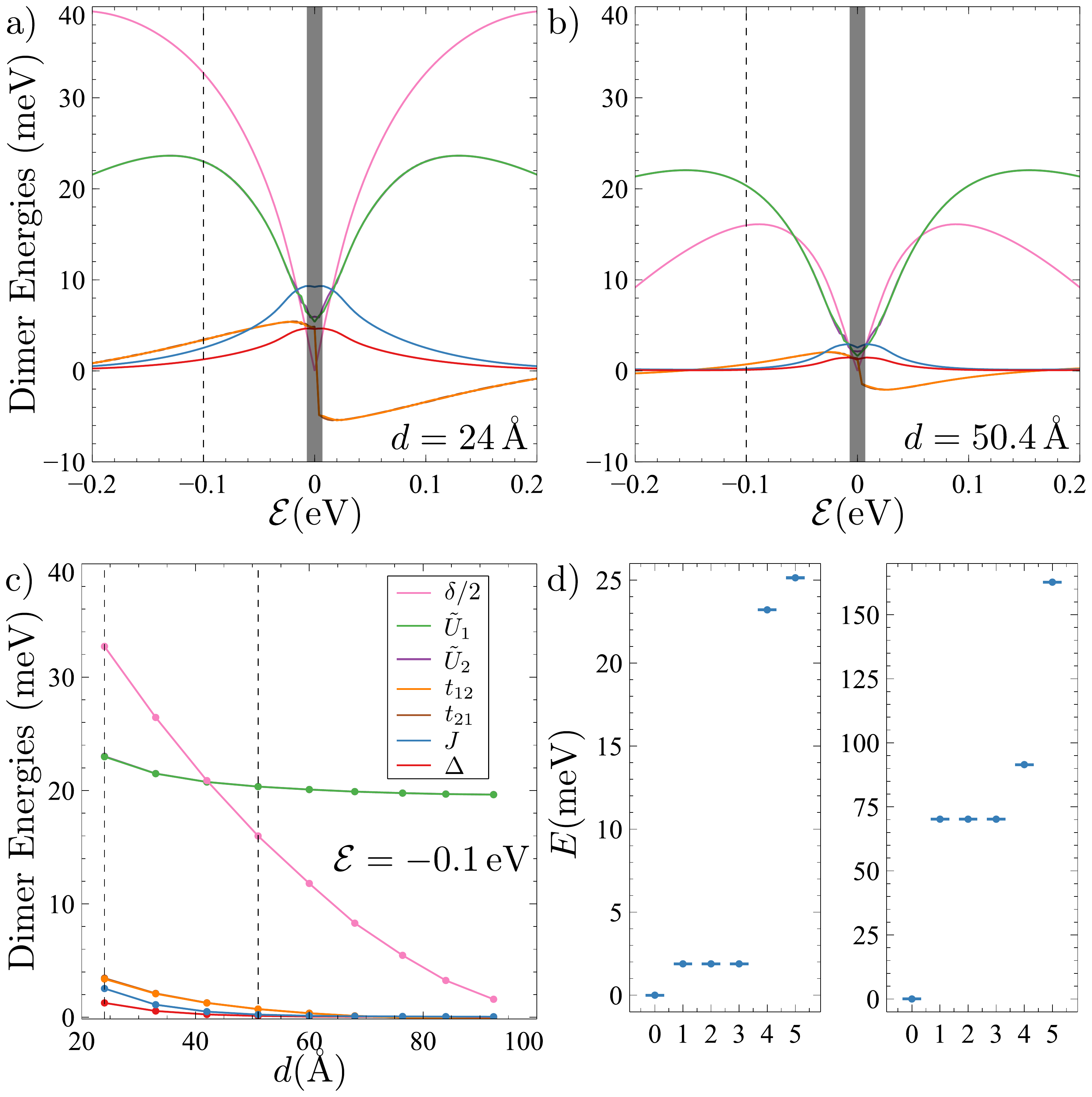}
\vspace{-5pt}
\caption{a,b) Dependence of $J$, $U_{1/2}$, $t_{12/21}$ and $\Delta$ on electric field for two sp$^3$ defects separated by $d=\SI{24}{\angstrom}$ and $\SI{50.4}{\angstrom}$ respectively. The dark stripes around $\mathcal{E}=0$ show the regime in which the confinement gap is comparable to the electric field induced gap. c) Dependence of the Hubbard parameters with the distance between sp$^3$ defects at $\mathcal{E}=\SI{-0.1}{\eV}$. d) Many body spectra at the strong ($d=\SI{92.4}{\angstrom}$, $\mathcal{E}=\SI{-0.2}{\eV}$) and weak coupling limit ($d=\SI{25.2}{\angstrom}$, $\mathcal{E}= \SI{-0.2}{\eV}$). The weak to strong coupling transition can be achieved changing either $d$ or $\mathcal{E}$.}
\label{blue_params}
\end{figure}

The evolution of each of these energy scales as we vary both the applied electric field and the dimer separation is shown in Fig.~\ref{blue_params} a-c). At zero field, the site states are sublattice polarized and the only finite energy scale is the ferromagnetic exchange $J$ and the pair hopping $\Delta$. As the field is ramped up, for both polarities, the site states acquire weight on both sublattices, which permits to have finite $\tilde{t}$, $\Delta$, and $t_{1,2}$. At the same time, their extension starts to shrink. The competition between these two effects leads both to the non monotonic behaviour of $\delta$, as well as to the decay of $J$ and $\Delta$. Importantly, for a sufficiently large electric field and defect separation, the dominant energy scales are $\tilde{t}$ and $\mathcal{U}$, showing that the artificial lattice defines an effective Hubbard model with narrow energy bands. 

We now discuss the properties of the multi-electronic eigenstates of $\mathcal{H}= \mathcal{H}_0+\mathcal{H}_U$.
We can expand these states in the basis of configurations with a well defined occupation of the single particle states $|\phi_{1,2}\rangle$. There are 6 such configurations for 2 electrons in 2 single particle states: four open-shell 
many-body states, where there is one electron per site-state and two closed shell states, where one of the site states is doubly occupied and the other is empty. The eigenstates of $\mathcal{H}$ have a well defined total spin $S$ that can take two values, $S=0$ and $S=1$. The triplet is formed with open shell configurations.
In contrast, the $S=0$ states combine both open and closed shell configurations.

For the dimer at half-filling, tuning the electric field can lead to three limiting cases with different types of ground state. For zero electric field, the only non-vanishing energy scales are $J$ and $\Delta$. Since the site state wave functions are real, we have $J=2\Delta$, leading to a $S=1$ ground state, complying with Lieb's theorem for a bipartite lattice\cite{Lieb1989}. 
At finite electric field both $\mathcal{U}$ and $\tilde{t}$
rapidly overcome the other energy scales, so that the system effectively becomes a Hubbard dimer. This leads to $S=0$ ground states. Here, we can have two different limiting situations\cite{Ortiz2018}. In the strong coupling limit, $\mathcal{U}\gg\tilde{t}\gg J$, the ground state is an open shell singlet with an entangled spin state. In the limit of weak coupling, $\tilde{t}\gg \mathcal{U}\gg J$, the system behaves like a Hydrogen molecule, with two electrons occupying a bonding molecular orbital. The corresponding many-body energy spectra for the weak and strong coupling cases at finite field are shown in Fig.~\ref{blue_params} d).

We now discuss qualitatively on the different electronic phases that could emerge in the limit of large lattices, on the basis of the predictions for a Hubbard model with first neighbour hoppings, although, other terms, such as beyond-first neighbour hopping, direct exchange and other many-body terms are also present. At half filling, the honeycomb and triangular lattices are expected to undergo a transition from a paramagnetic conducting phase to a Mott insulator that can either have magnetic order \cite{sorella1992} or be spin liquids \cite{Sahebsara08}, as the ratio $\tilde{U}/\tilde{t}$ goes above a critical threshold.

Away from half filling many other non-trivial phases have been predicted. For the triangular lattice at 3/4 filling, a non-coplanar spin canted phase with a Chern number $\mathcal{C}=2$ has been predicted\cite{martin08} giving rise to a quantized anomalous Hall effect. In the case of the honeycomb, chiral superconductivity has been predicted when the Fermi energy is tuned at the Van Hove singularities\cite{nandkishore2012}.
The application of perpendicular magnetic field would lead to the quantized regime at much smaller magnetic fields, on account of the larger unit cell. Finally, superconducting proximity effect, induced by lateral superconducting contacts, might bring an additional knob to induce novel electronic phases, including topological superconductivity.\cite{san2015}
Of course, the sp$^3$ functionalization approach could be used to create zero and one dimensional structures as well.
 
The experimental realization of this type of artificial lattices will face several technical challenges. So far, the manipulation of hydrogen atoms on graphene has been demonstrated with up to 14 atoms\cite{Brihuega2016}. In principle, there is no limit on the number of adatoms that can be manipulated\cite{khajetoorians2019}. For instance, up to 
80000 chlorine atoms on a Cu(111) surface were manipulated, making use of automatic protocols\cite{Kalff2016}.
Importantly, the number of localized states needed for emergent phases to occur
 is definitely smaller than that: the correlated phases observed in magic angle twisted bilayer graphene occur in samples with $10^4$ moiré unit cells.
A second important challenge will be the combination of STM manipulation and dual gating. A possible implementation would be to carry out the atomic manipulation on a graphene bilayer placed on a gated substrate and use the STM as a top gate. This would require to have electrodes on the sample to carry out in-plane transport to probe the electronic phases. 

In summary, we have analysed the potential of functionalized graphene bilayer to produce artificial fermion lattices where the band and Coulomb energies can be tuned by means of gate voltages and geometrical control of the defects. The system proposed here combines the advantage of electric control afforded by quantum dots arrays, and the advantage of reversible structural control given by STM atomic manipulation. The proposed platform could be used to explore a very large variety of electronic phases, including Mott-Hubbard transition, magnetically ordered phases, quantized anomalous insulators, spin liquids and superconducting phases.

\emph{Acknowledgements.-} 
We acknowledge J. L. Lado for fruitful discussions and technical assistance in the calculations.
We acknowledge grants P2020-PTDC/FIS-NAN/3668/2014, financial support Generalitat Valenciana funding Prometeo2017/139 and MINECO Spain (Grant No. MAT2016-78625-C2).

\bibliography{all_bib.bib}
\bibliographystyle{apsrev4-1}

\end{document}